\documentclass[12pt]{article}
\usepackage{graphicx}
\usepackage{epsfig}
\usepackage{latexsym}
\usepackage{latexsym}
\usepackage{amssymb}
\usepackage{amsmath}
\newcommand{\labell}[1]{\label{#1}}
\newcommand{\reef}[1]{(\ref{#1})}

\newcommand{\ud} {\mathrm{d}}

\def\Tr{{\rm Tr}}

\DeclareSymbolFont{AMSb}{U}{msb}{m}{n}
\DeclareMathSymbol{\IN}{\mathbin}{AMSb}{"4E}
\DeclareMathSymbol{\IZ}{\mathbin}{AMSb}{"5A}
\DeclareMathSymbol{\IR}{\mathbin}{AMSb}{"52}
\DeclareMathSymbol{\Q}{\mathbin}{AMSb}{"51}
\DeclareMathSymbol{\II}{\mathbin}{AMSb}{"49}
\DeclareMathSymbol{\IC}{\mathbin}{AMSb}{"43}
\DeclareMathSymbol{\IP}{\mathbin}{AMSb}{"50}
\DeclareMathSymbol{\IH}{\mathbin}{AMSb}{"48}
\DeclareMathSymbol\IA{\mathalpha}{AMSb}{"41}
\DeclareMathSymbol\IS{\mathalpha}{AMSb}{"53}

\def\Q{{\cal Q}}

\begin{document}

\begin{flushright}
%USC-04-05%\\
%DCPT--04/??
\end{flushright}
\begin{center} {\Large \bf B\"acklund Transformations, D--Branes,  and Fluxes}

\bigskip

{\Large\bf in}

\bigskip

{\Large\bf Minimal  Type 0 Strings}

\end{center}

\bigskip \bigskip \bigskip

\centerline{\bf James E. Carlisle${}^\flat$, Clifford V. Johnson${}^{\natural,}$\footnote{Also, Visiting Professor at the Centre for Particle Theory${}^\flat.$}, Jeffrey S. Pennington${}^\sharp$}

\bigskip
\bigskip

  \centerline{\it ${}^{\flat}$Centre for Particle Theory}
\centerline{\it Department of Mathematical
    Sciences}
\centerline{\it University of Durham}
\centerline{\it Durham DH1 3LE, England, U.K.}

\bigskip
\bigskip

  \centerline{\it ${}^{\natural,\sharp}$Department of Physics and Astronomy }
\centerline{\it University of
Southern California}
\centerline{\it Los Angeles, CA 90089-0484, U.S.A.}

\bigskip
\bigskip

\centerline{\small \tt j.e.carlisle@durham.ac.uk, johnson1@usc.edu, jspennin@usc.edu}

\bigskip
\bigskip

%\maketitle

\begin{abstract}

  We study the Type 0A string theory in the $(2,4k)$ superconformal
  minimal model backgrounds, focusing on the fully non--perturbative
  string equations which define the partition function of the model.
  The equations admit a parameter, $\Gamma$, which in the spacetime
  interpretation controls the number of background D--branes, or R--R
  flux units, depending upon which weak coupling regime is taken. We
  study the properties of the string equations (often focusing on the
  $(2,4)$ model in particular) and their physical solutions. The
  solutions are the potential for an associated Schr\"odinger problem
  whose wavefunction is that of an extended D--brane probe.  We
  perform a numerical study of the spectrum of this system for varying
  $\Gamma$ and establish that when $\Gamma$ is a positive integer the
  equations' solutions have special properties consistent with the
  spacetime interpretation.  We also show that a natural
  solution--generating transformation (that changes $\Gamma$ by an
  integer) is the B\"acklund transformation of the KdV hierarchy
  specialized to (scale invariant) solitons at zero velocity. Our
  results suggest that the localized D--branes of the minimal string
  theories are directly related to the solitons of the KdV hierarchy.
  Further, we observe an interesting transition when $\Gamma=-1$.

\end{abstract}
\newpage \baselineskip=18pt \setcounter{footnote}{0}

%%%%%%%%%%%%%%%%%%%%%%%%%%%%%%%%%%%%%%%%%%%%%%%%%%%%%%%%%%%%%%%%%%%%%%%%%%%%%%%
\section{Background}
\label{sec:introduction}

The $(2,4k)$ series of the minimal type 0A string theory, in the
presence of background R--R sources, has a non--perturbative
definition {\it via} the following ``string
equation''\cite{Dalley:1992br}:
\begin{equation}
u{\cal R}^2-\frac{1}{2}{\cal R}{\cal R}^{''}+\frac{1}{4}({\cal
R}^{'})^2
  =\nu^2\Gamma^2\ .\labell{eq:nonpert}
\end{equation}

Here $u(z)$ is a real function of the real variable $z$; a prime
denotes $\nu \partial/\partial z$; and $\Gamma$ and $\nu$ are
constants. The quantity $\mathcal{R}$ is defined by:
\begin{equation}
\label{eq:R} {\cal
  R}=\sum_{k=0}^\infty  \left(k+\frac{1}{2}\right)t_k R_k\ ,
\end{equation}
where the $R_k$ ($k=0,\ldots$)
 are polynomials in $u(z)$ and its $z$--derivatives. They
are related by a recursion relation:
\begin{equation}
  \labell{eq:recursion}
  R^{'}_{k+1}=\frac{1}{4}R^{'''}_k-uR^{'}_k-\frac{1}{2}u^{\prime}R_k\ ,
\end{equation}
and are fixed by the constant $R_0$, and the requirement that the rest
vanish for vanishing $u$. The first few are:
\begin{equation}
  \labell{eq:firstfew}
  R_0=\frac{1}{2}\ ;\quad R_1=-\frac{1}{4}u\ ;\quad R_2=\frac{1}{16}(3u^2-u^{''})\ .
\end{equation}
The $k$th model is chosen by setting all the other $t \,$s to zero
except $t_0\equiv z$, and $t_k$, the latter being fixed to a
numerical value such that ${\cal R}={\cal D}_k-z$. The ${\cal
D}_k$ are normalised such that the coefficient of $u^k$ is unity,
{\it e.g.}:
\begin{eqnarray}
{\cal D}_1=u\ ,\quad {\cal D}_2=u^2-\frac{1}{3}u^{''}\ ,\quad {\cal D}_3=u^3- u u^{''}-\frac{1}{2}(u^{'})^2+\frac{1}{10}u^{''''}\ .
  \labell{eq:diffpolys}
\end{eqnarray}

For the $k$th model, equation~\reef{eq:nonpert} has asymptotics:
\begin{eqnarray}
  u(z)&=&z^{\frac{1}{k}}+\frac{\nu\Gamma}{kz^{1-\frac{1}{2k}}}+\cdots\quad \mbox{\rm for}\quad z\longrightarrow +\infty\ ,
\nonumber\\
u(z)&=&\frac{\nu^2(4\Gamma^2-1)}{4z^2}+\cdots\quad \mbox{\rm for}\quad z\longrightarrow -\infty\ .
  \label{eq:largez}
\end{eqnarray}

The function $u(z)$ defines the partition function $Z=\exp(-F)$ of the
string theory {\it via}:
\begin{equation}
u(z)=\nu^2\frac{\partial^2 F}{\partial \mu^2}\Biggl|_{\mu=z}\ ,
  \labell{eq:partfun}
\end{equation}
where $\mu$ is the coefficient of the lowest dimension operator in
the world--sheet theory. Integrating twice, the asymptotic
expansions in equations~\reef{eq:largez} furnish the partition function
perturbatively as an expansion in the dimensionless string
coupling
\begin{equation}
  \label{eq:stringcoupling}
g_s={\nu\over\mu^{1+{1\over2k}}}\ .
\end{equation}
The role of $\Gamma$ is now clear: at large positive $z$ we find
that $\Gamma$ controls the number of background D-branes since the
perturbative series contains both open and closed string
worldsheets with a factor of $\Gamma$ for each boundary; while the
large negative $z$ series gives only closed string worldsheets,
with $\Gamma^2$ appearing when there is an insertion of pure R--R
flux\cite{Dalley:1992br,Klebanov:2003wg}.

From the point of view of the $k$th theory, the other $t_k$s
represent coupling to closed string operators ${\cal O}_k$. It is
well known that the insertion of each operator can be expressed in
terms of the KdV flows\cite{Douglas:1990dd,Banks:1990df}:
\begin{equation}
  \labell{eq:kdvflows}
  \frac{\partial u}{\partial t_k}= R^{'}_{k+1}\ .
\end{equation}
The operator ${\cal O}_0$ couples to $t_0$, which is in fact $-4z$,
the cosmological constant (in the unitary model). So ${\cal O}_0$ is
often referred to as the puncture operator, which yields the area of a
surface by fixing a point which is then integrated over in the path
integral. So $u(z)$ is the two--point function of the puncture
operator.

For $\Gamma=0$ the string equation \reef{eq:nonpert} was
discovered by defining a family of string theories using double
scaled models of a complex matrix $M$:
\begin{equation}
{\cal Z}=\int dM \exp\left\{-\frac{N}{\gamma}{\rm \Tr}\left(V(M
M^\dagger)\right)\right\}\ .
  \labell{eq:complex}
\end{equation}
The resulting physics captured by the string equation was the first
complete non--perturbative definition of a string
theory\cite{Dalley:1992qg,Dalley:1992vr,Dalley:1992yi,Johnson:1992pu,Johnson:thesis},
sharing the large $z$ perturbation theory of the original (bosonic)
string theories obtained by double scaling of Hermitian matrix
models\cite{Brezin:1990rb,Douglas:1990ve,Gross:1990vs}, but not
suffering from their non--perturbative shortcomings.

Non--zero $\Gamma$ can also be studied using a matrix model definition
in a variety of ways. One way is to add a logarithmic
term\cite{Kazakov:1990cq,Kostov:1990nf} to an
appropriate\cite{Dalley:1992br,Johnson:2004ut} matrix model potential.
Expanding the logarithm clearly adds holes of all sizes to the string
worldsheets defined by studying the duals of the Feynman diagrams of
the model.  Another (equivalent) method is to define $M$ as an
$(N+\Gamma) \times N$ rectangular
matrix\cite{Lafrance:1993wy,Klebanov:2003wg}. The double scaling limit
(in which $N$ is taken large) then yields equation \reef{eq:nonpert}.

It is clear from these two methods that, in the interpretation of the
matrix model as the world--volume theory of $N$ D--branes (which
defines the closed string theory holographically\cite{McGreevy:2003kb}
after taking the double scaling limit), the modification corresponds
to adding $\Gamma$ ``quark flavours'' to the world--volume model. This
corresponds to adding sectors of open strings stretching between the
$N$ D--branes and $\Gamma$ extra D--branes.

So from this point of view it appears that $\Gamma$ is a positive
integer. However, this is not at all clear from the string equation
itself. In fact, it is known perturbatively that the solutions of the
equation have special properties for various fractional values of
$\Gamma$ \footnote{Note in particular the case $\Gamma=\pm1/2$ in the
  large negative $z$ regime of equation~\reef{eq:largez}. There are
  several other interesting cases of this type, including a family of
  exact double pole solutions, for half--integer $\Gamma$. See
  refs.\cite{Dalley:1992br,Johnson:2004ut}.}. These might well turn
out to be unphysical values, but this is not {\it \`a priori} clear.
One of the purposes of this paper is to show that $\Gamma$ being a
positive integer is in fact selected out as special from the point of
view of the string equation. We will do this by studying the equation
non--perturbatively using numerical methods, and by also observing the
properties of a natural transformation between solutions of the string
equation which change $\Gamma$ by $\pm1$. This transformation was
discovered in ref\cite{Dalley:1992br}, and we derive an explicit form
for it here, and employ this form in our investigations. We further
observe that the transformation is in fact a specialisation (to the
scale invariant soliton sector) of the B\"acklund transformations of
the KdV hierarchy!

These B\"acklund transformations are known to change the soliton
number of solutions of the KdV equation (and its higher order
analogues)\cite{Das:1989fn}, and so our observation is
particularly intriguing, since it implies that the
(localized\cite{Zamolodchikov:2001ah}) D--branes of the minimal
type~0A string theory {\it are intimately connected with the
solitons of the KdV hierarchy}.

\bigskip

{\bf Note:} While this paper was being written, a paper
appeared\cite{Seiberg:2004ei} which contains results that have some
overlap with ours.

\section{A Spectral Problem}

A solution $u(z)$ of the string equation serves as the potential for a
one--dimensional Hamiltonian ${\cal H}$ which arises naturally in the
double--scaled matrix model\cite{Gross:1990aw}:
\begin{equation}
{\cal H}=-Q=-\nu^2\frac{\partial^2}{\partial z^2}+u(z)\ ,
  \label{eq:hamiltonian}
\end{equation} where $\sqrt{2}\,\nu$ plays the role of $\hbar$.

In fact, the (first derivative of) the string equation can be obtained
by eliminating the wave--function $\psi(z)$ from the equations:
\begin{eqnarray}
Q\psi&=&\lambda_s\psi\ ,\nonumber\\
{\widetilde P}\psi&=&\lambda_s{\partial\over\partial\lambda_s}\psi\ .
  \label{eq:PQQ}
\end{eqnarray}
Here, ${\widetilde P}$ is a differential operator representing scale
transformations. It is constructed from (the differential operator
part of) fractional powers\cite{Gelfand:1976A,Gelfand:1976B} of $Q$,
as follows\cite{Dalley:1992vr}:
\begin{equation}
{\widetilde P}=\sum_{k=1}^{\infty}\left(k+\frac12\right)t_k Q_+^{k+\frac12}-\frac12 z\frac{\partial}{\partial z}+{\rm const.}\ .
  \label{eq:ptilde}
\end{equation}
Eliminating $\psi$, one has the operator
equation\cite{Dalley:1992vr}:
\begin{equation}
  \label{eq:pqq}
  [{\widetilde P},Q]=Q\ ,
\end{equation}
which yields an equation for $u(z)$ which is the first derivative of
equation \reef{eq:nonpert}.

The spectrum of this model has interesting and important information,
and is a rather direct way of getting access to the non--perturbative
physics of the model\footnote{The wavefunction $\psi$ is a natural
  extended D--brane\cite{Fateev:2000ik} probe of the model, a fact
  which has been studied extensively recently for the bosonic string
  in ref.\cite{Maldacena:2004sn}.}. \ The problem is naturally defined
in terms of the eigenvalues, $\lambda$, of the combination
$ MM^\dagger$.  These are naturally positive. So in the double
scaling limit (which focuses on the infinitesimal region near the
tail of the eigenvalue distribution of the model) the scaled
eigenvalues, which we denote as $\lambda_s$, are also distributed
on the positive real line.

We verified this in the full non--perturbative regime by studying
the spectrum of ${\cal H}$ numerically. To do this we solved the
differential equation for a given value of $k$ as a boundary value
problem, with the perturbative boundary conditions given in
equations~\reef{eq:largez}. We started out by using the
equation--solving routine {\tt dsolve} in the package {\it Maple
9}, although later we used other methods (see below for the case
of negative $\Gamma$). We used a discrete $z$ lattice of up to
$8192$ points for positive $\Gamma$.

\begin{figure}[ht]
\begin{center}
\includegraphics[scale=0.55]{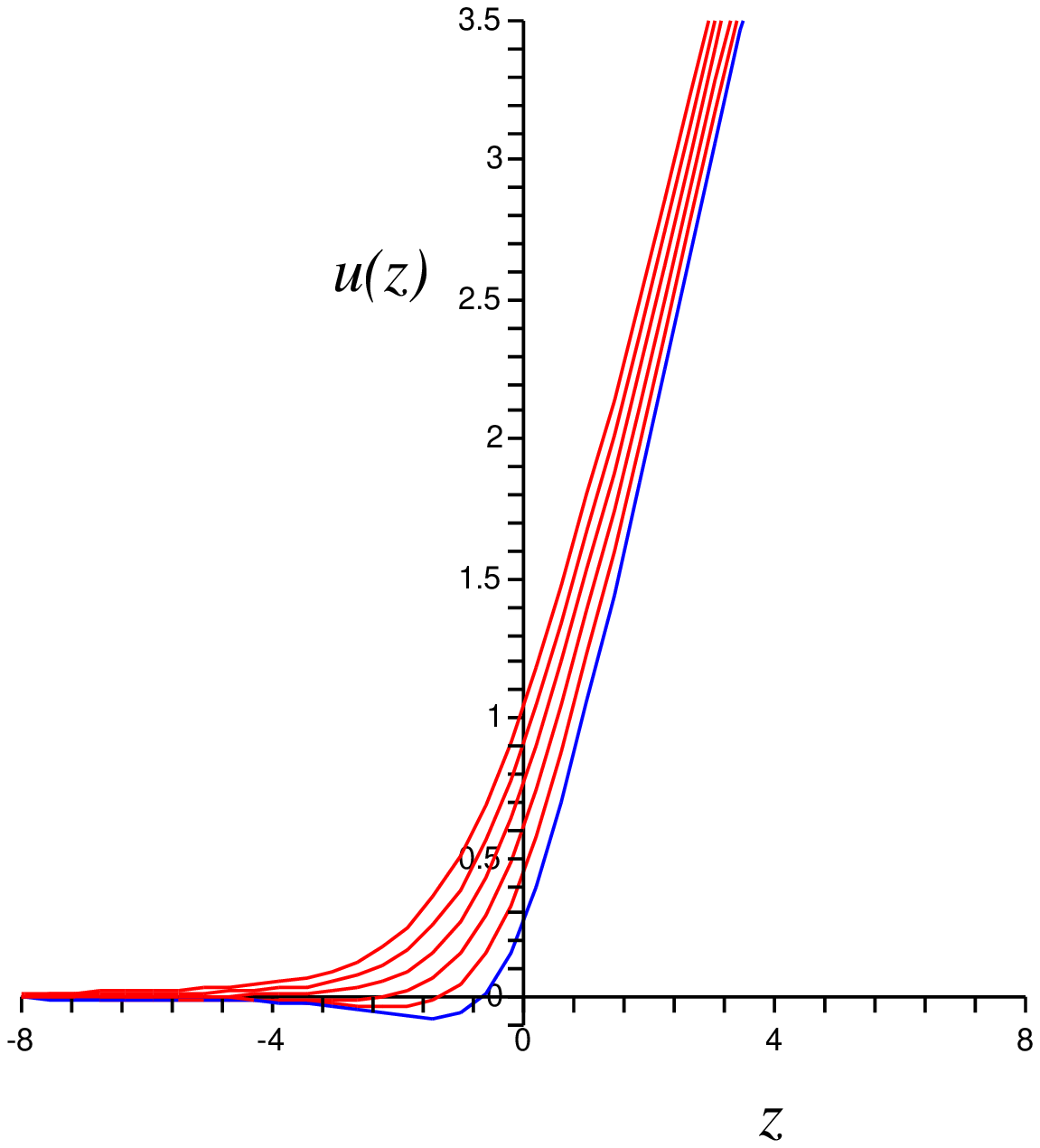} \includegraphics[scale=0.55]{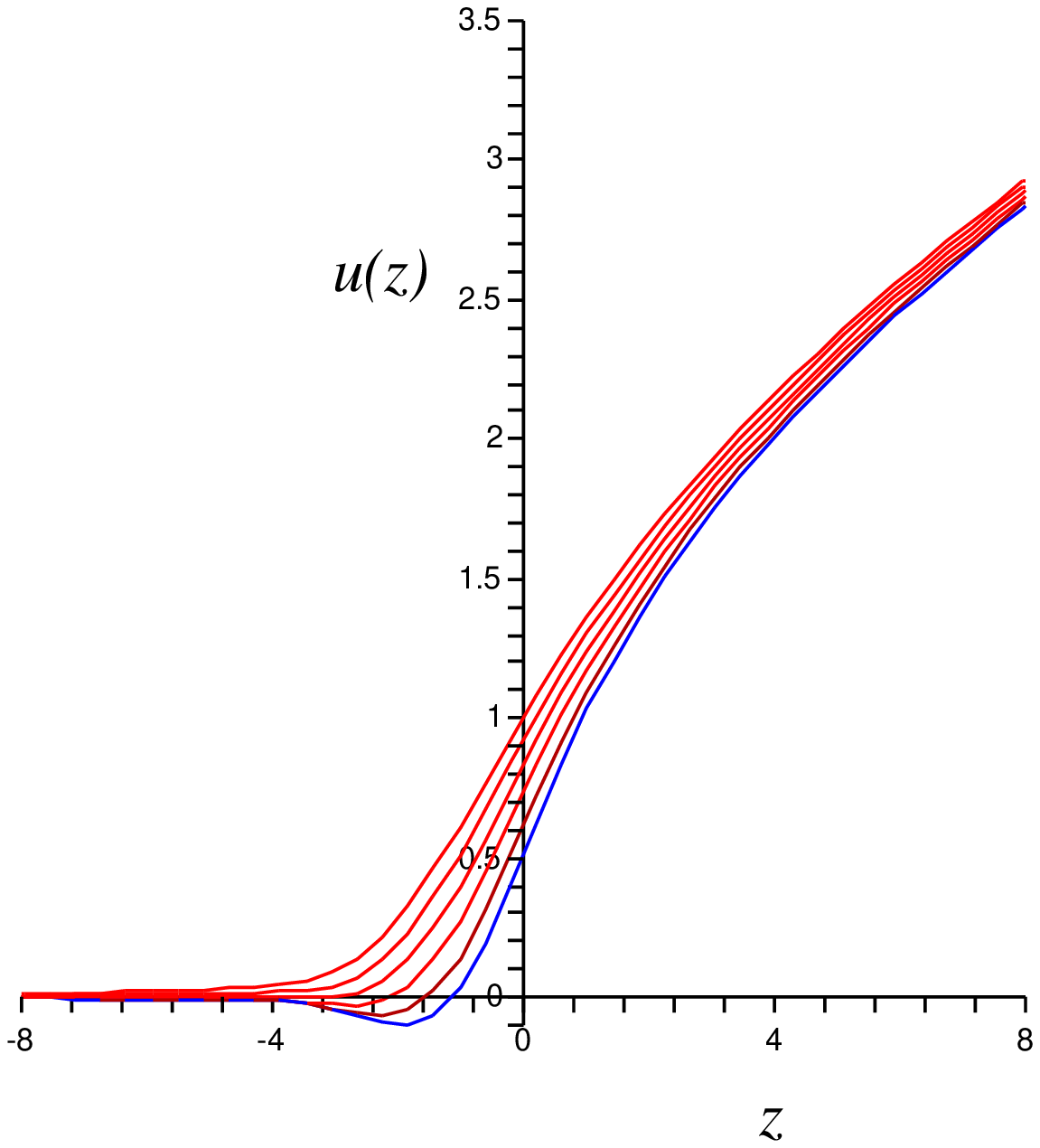}
\end{center}
\caption{\small Numerical solutions to equation~\reef{eq:nonpert}
for $u(z)$: {\it (a)} The case $k=1$, and {\it (b)} the case
$k=2$, for a range of values of  the parameter $\Gamma$. The very
bottom curve in each case has  $\Gamma=0$. The solutions
all asymptote to $u=z^{1/k}$ for large positive $z$ and $u=0$ for
large negative $z$.} \label{fig:gammaplots}
\end{figure}

First, note the form of the function $u(z)$ for some {\it strictly
positive} values of $\Gamma$ (see figure~\ref{fig:gammaplots}).
Since the potential tends to zero at $z\to-\infty$, and rises
monotonically in the $z\to+\infty$ limit, it is clear that the
spectrum has the chance to be both continuous and bounded from below
by zero, as we expect from perturbation theory. A striking feature is
the fact that the potential develops a potential well in the interior,
as $\Gamma\to0$.

It is therefore interesting and important to {\it non--perturbatively}
establish these properties (continuity, boundedness) of the spectrum
for various $\Gamma$. First, we checked the spectrum of the case
$\Gamma=0$ (the case with no background branes), and verified that
there are no bound states, despite the appearance of the
well\footnote{That this is the case exactly fits with a suggestive
  (but rough) estimate one can make by comparing the depth of the well
  with the value of the lowest energy eigenstate of the equivalent
  harmonic oscillator, determined by reading off the value of the
  second derivative at the bottom of the well (computed numerically).
  The well falls short in depth by $\sim10\%$.}\footnote{This
  complements earlier studies of the properties of the $\Gamma=0$
  solutions carried out extensively in
  refs.\cite{Dalley:1992qg,Dalley:1992vr,Dalley:1992yi,Johnson:1992pu,Johnson:thesis}.}.
We set up the problem numerically by discretizing the spatial
coordinate $z$, typically using the interval $-10\leq z\leq +10$, with
$\nu=1$. We used a range of lattice spacings in studying our problem
to ensure stability. This was equivalent to a number of lattice points
ranging from as little as 500 up to 10000 (the latter used when more
accuracy was needed, as we shall see later on.)  Discretizing this
problem turns it into a problem of diagonalizing a tri--diagonal
matrix for which we coded a fairly efficient well--known method (the
TQLI method\cite{recipes}) in C$++$ to do the work.  This allowed us
to establish non--perturbatively with some confidence that there are
no bound states in the well for positive $\Gamma$, and so the spectrum
is continuous and bounded from below by zero, in accord with
perturbative expectations.

\begin{figure}[ht]
\begin{center}
\includegraphics[scale=0.6]{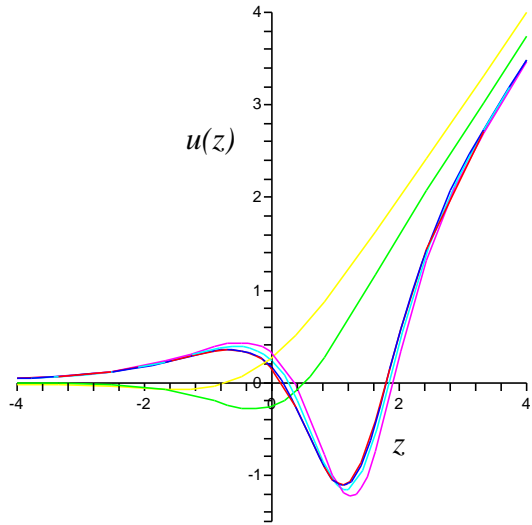}
\end{center}
\caption{\small Numerical solutions to equation~\reef{eq:nonpert} for $u(z)$ in the case $k=1$, for a range of negative values of  the parameter $\Gamma$:  $0.000,-0.500,-0.950,-0.965,-0.968,-0.970$.}
\label{fig:gammanegplots}
\end{figure}

The next step is the case of negative $\Gamma$, which is clearly
allowed by the equation and so should be studied. Our first
observation is that solving the differential equation numerically
became much more prone to error, and we had to use much more accurate
methods. We used the NAG library of routines for solving boundary
value problems with C$++$ to make further progress, as they gave much
greater control over the problem.  We were able to construct directly
in this way a solution for $u(z)$ for values as low as $\Gamma=-0.970$
before running into numerical problems. See
figure~\ref{fig:gammanegplots}.

Once again we should check that the perturbative expectation that the
spectrum is positive persists non--perturbatively. Several solutions
$u(z)$, for various $\Gamma$ down to $-0.970$ were used as potentials
for ${\cal H}$ and found (upon placing in our tridiagonal matrix
diagonalization routine) to have no bound states.

With these methods, the growing instability is suggestive, but not
compellingly so, that something interesting might be happening at
$\Gamma=-1$. In order to establish this more satisfyingly we employed
some exact methods (in the form of a solution--generating
transformation), which we will describe next. This will allow us to
generate new $u(z)$ from the ones we already can extract numerically,
and allow us to approach $\Gamma=-1$ in a controlled manner, and with
convincing accuracy.

\section{B\"acklund Transformations}
In ref.\cite{Dalley:1992br}, it was established that the double scaled
unitary matrix models of refs.\cite{Periwal:1990gf,Periwal:1990qb} had
an interpretation as continuum string theories with open string
sectors. This was done by establishing a direct connection between the
solutions of the string equations in equation~\reef{eq:nonpert} and
solutions of the string equations for those unitary matrix models,
which are most naturally written in terms of the quantities $S_k$,
where:
\begin{eqnarray} \labell{eq:DJMW-Poly}
S_k \equiv \frac12 R_k^\prime [v^2 + v^\prime] - v R_k[v^2 +
v^\prime]\ ,
\end{eqnarray}
as:
\begin{eqnarray} \labell{eq:DJMW-5}
\sum_{k=1}^{\infty} \left(k + \frac12\right) t_k S_k [v(z)] + z v(z) = \nu C\ .
\end{eqnarray}
The models are organised by the mKdV hierarchy:
\begin{eqnarray} \labell{eq:DJMW-Flow}
\frac{\partial v}{\partial t_k} = \frac12  S^\prime_{k}[v]\ .
\end{eqnarray}
We define $u(z)$ and $v(z)$ such that:
\begin{eqnarray} \labell{eq:DJMW-7}
X_{\pm} [u, v] \equiv \frac12 \mathcal{R}^\prime [u] \mp \nu\Gamma -
v(z)\mathcal{R}[u] = 0\ ,
\end{eqnarray}
which implies a specific form for $v(z)$ given a $u(z)$:
\begin{eqnarray} \labell{eq:DJMW-8}
v = \frac{\frac12 \mathcal{R}^\prime [u] \mp \nu\Gamma
}{\mathcal{R}[u]}\ .
\end{eqnarray}
Noting the identity\cite{Dalley:1992br}:
\begin{eqnarray} \labell{eq:DJMW-9}
0 = X_{\pm}^2 \pm \nu\Gamma X_{\pm} - \mathcal{R}[u] X_{\pm}^\prime
\equiv (v^2 + v^\prime) \mathcal{R}^2 [u] - \frac12 \mathcal{R}[u]
\mathcal{R}^{\prime \prime}[u] + \frac14 (\mathcal{R}^\prime
[u])^2 - \nu^2\Gamma^2\ ,
\end{eqnarray}
we see that if the inverse transformation $u=v^2 + v^\prime$
(known in the integrable literature as the Miura map) exists, then
this is just our original string equation~\reef{eq:nonpert}; and
on substitution into equation~\reef{eq:DJMW-8} gives
equation~\reef{eq:DJMW-5} with $C=1/2 \pm \Gamma$.  Since the
unitary matrix models of refs.\cite{Periwal:1990gf,Periwal:1990qb}
were originally derived with $C=0$, those models turn out to be
identified with the case $\Gamma=1/2$.

Since the transformation is true for both values of $C$ we find that a
given $u(z)$ for a value of $\Gamma$ gives two functions $v_C(z)$ and
$v_{1-C}(z)$, which must be related by:
\begin{eqnarray} \labell{eq:DJMW-10a}
v_C^2 + v_C^\prime = v_{1-C}^2 + v_{1-C}^\prime\ ,
\end{eqnarray}
and hence:
\begin{eqnarray} \labell{eq:DJMW-10b}
v_{1-C} = v_C + \frac{2C-1}{\mathcal{R} [v_C^2 + v_C^\prime]}\ .
\end{eqnarray}
On the other hand, the string equation~\reef{eq:DJMW-5} has the
symmetry $v_{-C}= -v_{C}$, and so we obtain a result which relates
$v(z)$ at one value of $C$ to another for $C\pm1$:
\begin{eqnarray} \labell{eq:DJMW-12}
v_{C \pm 1} =- v_C - \frac{2C \pm 1}{\mathcal{R} [v_C^2 \mp
v_C^\prime]}\ .
\end{eqnarray}
This implies a similar transformation for functions $u(z)$, relating $\Gamma$
and $\Gamma\pm1$. We can make this explicit by combining
equations~\reef{eq:DJMW-8}, \reef{eq:DJMW-9}, \reef{eq:DJMW-12} and $u
= v^2 + v^\prime$ to give the remarkable result:
\begin{eqnarray} \labell{eq:Back-Explicit-Gam}
u_{\Gamma \pm 1} = \frac{3 \left(\mathcal{R}^{\prime} \right)^2 -
2 \mathcal{R}\mathcal{R}^{\prime \prime} \mp 8 \nu\Gamma \,
\mathcal{R}^{\prime} + 4 \nu^2\Gamma^2}{4 \mathcal{R}^2}\ ,
\end{eqnarray}
where $\mathcal{R} \equiv \mathcal{R}(u_{\Gamma})$.

In the theory of non--linear differential equations (such as those
encountered in the integrable model context) B\"acklund
transformations are extremely useful, since they convert a given
solution of the differential equation into a new
solution\footnote{These are actually ``auto--B\"acklund''
  transformations.}. Strictly speaking, it looks rather like this is
{\it not} what we have here since our string
equation~\reef{eq:nonpert} is displayed with $\Gamma$ explicitly
appearing, so $u_\Gamma$ and $u_{\Gamma\pm1}$ are solutions of {\sl
  different} equations. However, the once--differentiated string
equation (which is in fact the one which appears naturally in many
derivations; see the beginning of the previous section, and see below), which is
\begin{equation}
  \label{eq:nonpertdiff}
  \frac12 {\cal R}^{\prime\prime\prime}-2u {\cal R}^\prime-u^\prime{\cal R}=0\ ,
\end{equation}
does not have an explicit appearance of $\Gamma$. The solutions
$u_\Gamma$ do know of $\Gamma$, of course, since it appears (for
example) in their asymptotic expansion~\reef{eq:largez}.

So we have a genuine B\"acklund transformation for our system. What is
particularly interesting is that it is not just any such
transformation but a special case of the well--known B\"acklund
transformations known to change the number of solitons in the KdV
hierarchy. We can make this explicit as follows:

We start with the KdV equations~\reef{eq:kdvflows}, which we rewrite
as:
\begin{eqnarray} \labell{eq:KdV-Gen}
\frac{\partial {\tilde u}}{\partial t_k} =  R^\prime_{k+1}[{\tilde u}] = \left(
\frac{1}{4} \ud^3 -  {\tilde u} \ud - \frac{1}{2} \frac{\partial
{\tilde u}}{\partial x} \right) R_k[{\tilde u}]\ ,
\end{eqnarray}
 where now $d \equiv \frac{\partial}{\partial x}$.

 We can search for solutions of the form\footnote{More generally we
   search for solutions of the form ${\tilde u}(x,t_k) =
   t_k^{\alpha_k} u(x t_k^{\beta_k})$; but we find that consistency
   forces us to choose $\alpha_k = 2 \beta_k$.} ${\tilde u}(x,t_k) =
 t_k^{2 \beta_k} u(x t_k^{\beta_k})$. We then find $\beta_k =
 -1/(2k+1)$ and that (under appropriate rescaling of $t_k$)
 equation~\reef{eq:KdV-Gen} becomes:
\begin{eqnarray} \labell{eq:String-1}
-u - \frac{1}{2} z u^\prime = \left( \frac{1}{4} \ud^3 -  u \ud -
\frac{1}{2} u^\prime \right) R_k[u]\ ,
\end{eqnarray}
where a prime denotes differentiation with respect to $z = x
t^{\beta_k}$. Re--arranging we find the once--differentiated
string equation~\reef{eq:nonpertdiff} for a particular $k$.

This is another way of making the observation (already noted in
ref.\cite{Dalley:1992vr}) that the string equation follows a
restriction of the KdV hierarchy to scale invariant solutions.  This
also fits with the recovery\cite{Dalley:1992vr} of the string equation
from the operator relation $[{\widetilde P},Q]=Q$, where ${\widetilde
  P}$ is the generator of scale transformations\footnote{The
  $[{\widetilde P},Q]=Q$ formalism of ref.\cite{Dalley:1992vr} allows
  any arbitrary combination of the $t_k$s to be switched on at once.}.
(See the previous section.)

Now note that the KdV hierarchy admits the following well known
B\"acklund transformation relating a solution ${\tilde u}_1$ to a solution
${\tilde u}_2$:
\begin{eqnarray} \labell{eq:Back-With-Lambda}
w_x + y_x = \frac{1}{2}(w - y)^2 + 2 {\tilde \lambda}\ ,
\end{eqnarray}
where $w_x = {\tilde u}_1$ and $y_x = {\tilde u}_2$.

This can be derived from the (generalised) Miura map ${\tilde
  u}=v^2+v^\prime + {\tilde \lambda}$ (where ${\tilde \lambda}$ is a
constant), which relates the KdV and mKdV hierarchies, by noting that
the mKdV flow hierarchy (see equation~\reef{eq:DJMW-Flow}) is
invariant under $v \rightarrow -v$. So we can have two solutions of
KdV, ${\tilde u}_1$ and ${\tilde u}_2$, arising from the same
solution, $v$, of the mKdV hierarchy.

 We have ${\tilde u}_1= w_x = v^2 + v^\prime +{\tilde  \lambda}$ and ${\tilde
   u}_2= y_x = v^2 - v^\prime + {\tilde \lambda}$, which gives upon addition
 and subtraction:
\begin{eqnarray} \labell{eq:Back-Deriv-1}
w_x + y_x = 2 v^2 + 2 {\tilde \lambda} \ ,\quad
w_x - y_x = 2 v^\prime\ .
\end{eqnarray}
The latter can then be integrated once and substituted into the
former to give equation~\reef{eq:Back-With-Lambda}. This can then
be specialized to solutions of the string equation by writing
${\tilde u}_i = t^{2 \beta_k} u_i(x t^{\beta_k}) \equiv t^{2
\beta_k} f_i^\prime(x t^{\beta_k})$, which gives:
\begin{equation}
  \label{eq:given}
 w = t^{\beta_k} f_1(x
t^{\beta_k})\ ,\qquad  y = t^{\beta_k} f_2(x t^{\beta_k})\ ,
\end{equation}
where a prime denotes differentiation with respect to $z = x
t^{\beta_k}$.  We have therefore
\begin{eqnarray} \labell{eq:Back-Scaled}
f_1^{\prime} + f_2^{\prime} = \frac{1}{2}(f_1 - f_2)^2 + 2 {\tilde
\lambda} t^{-2 \beta_k}\ .
\end{eqnarray}
So we see that for consistency we must set ${\tilde\lambda}=0$, which
has an interesting interpretation to be discussed below. Hence we have
\begin{eqnarray} \labell{eq:Back-Without-Lambda}
f_1^\prime + f_2^\prime = \frac{1}{2}(f_1 - f_2)^2\ .
\end{eqnarray}
So far, it is not clear that this transformation changes $\Gamma$. In
order to establish the connection between this transformation and the
one displayed in equation~\reef{eq:Back-Explicit-Gam}, in which
$\Gamma$ appears explicitly, we first work perturbatively.  Starting
with the asymptotic expansions~\reef{eq:largez}, for (say) $u_1$, it
is easy to show that using equation~\reef{eq:Back-Without-Lambda} the
asymptotic expansion obtained for $u_2$ is of the same form, except
that $\Gamma$ has been replaced by $\Gamma\pm1$. We then expect that
they are equivalent non--perturbatively, and have checked that this is
the case by working numerically on some explicit solutions.

So we have established that the B\"acklund transformation takes us
between solutions with asymptotics given in equation~\reef{eq:largez}
which are for $\Gamma$s differing from each other by unity.  It is
known\cite{Das:1989fn} that the B\"acklund
transformations~\reef{eq:Back-With-Lambda} of KdV increase or decrease
the soliton number of a given solution by unity.  In fact, for a given
solution $u$, the soliton corresponds to a bound state of ${\cal H}$
with (negative) eigenvalue ${\tilde\lambda}$, and the speed of the
soliton is given by $-{\tilde\lambda}$ (up to a positive numerical
constant).  The soliton number of a particular solution can be read
off by the number of such bound states and their individual velocities
are set by the discrete spectrum.

In our case, we have established (so far) that there are no bound
states in our solutions, and indeed, our restriction of the B\"acklund
transformation forces ${\tilde\lambda} =0$. So we deduce from this
that $\Gamma$, which counts soliton number, must count the number of
{\it zero--velocity} solitons. This is another sign that $\Gamma$ is
naturally integer, and positive\footnote{It is amusing to note that
  one can generate non--trivial solutions of the string equation by
  starting with the trivial solution $u=0$. This does not give
  solutions with the ``physical'' asymptotics given in
  equation~\reef{eq:largez}, which is not a contradiction, since we
  did not start with such a solution. What one obtains are pole
  solutions with the half--integer values of $\Gamma$ similar to those
  discussed in refs.\cite{Dalley:1992br,Johnson:2004ut}.}. We will
shortly find further evidence for this.

Now that we have a method for generating new solutions $u(z)$ starting
from old ones, we can return to our numerical study armed with more
powerful tools.

\section{The Case of $\Gamma=-1$}

We established in the last section that given a solution $u_\Gamma$,
we can generate a solution $u_{\Gamma\pm1}$. In the section before, we
reported that it was difficult to solve the string equation
numerically as $\Gamma$ became more negative.  Numerical precision was
lost rapidly as $\Gamma$ went below the value $-0.9$, while positive
$\Gamma$ is very much under control.  We can use our transformation
from the previous section to surmount this obstacle, by simply solving
the equation numerically for $u(z)$ for $\Gamma=\epsilon$, for
$\epsilon$ small and positive, and then use our
transformation~\reef{eq:Back-Explicit-Gam} to build a solution for
$u(z)$ with $\Gamma=-1+\epsilon$. We can simply follow $\epsilon$ to
vanishingly small values and therefore learn the properties of $u(z)$
on the approach to $\Gamma=-1$.

\begin{figure}[ht]
\begin{center}
\includegraphics[scale=0.55]{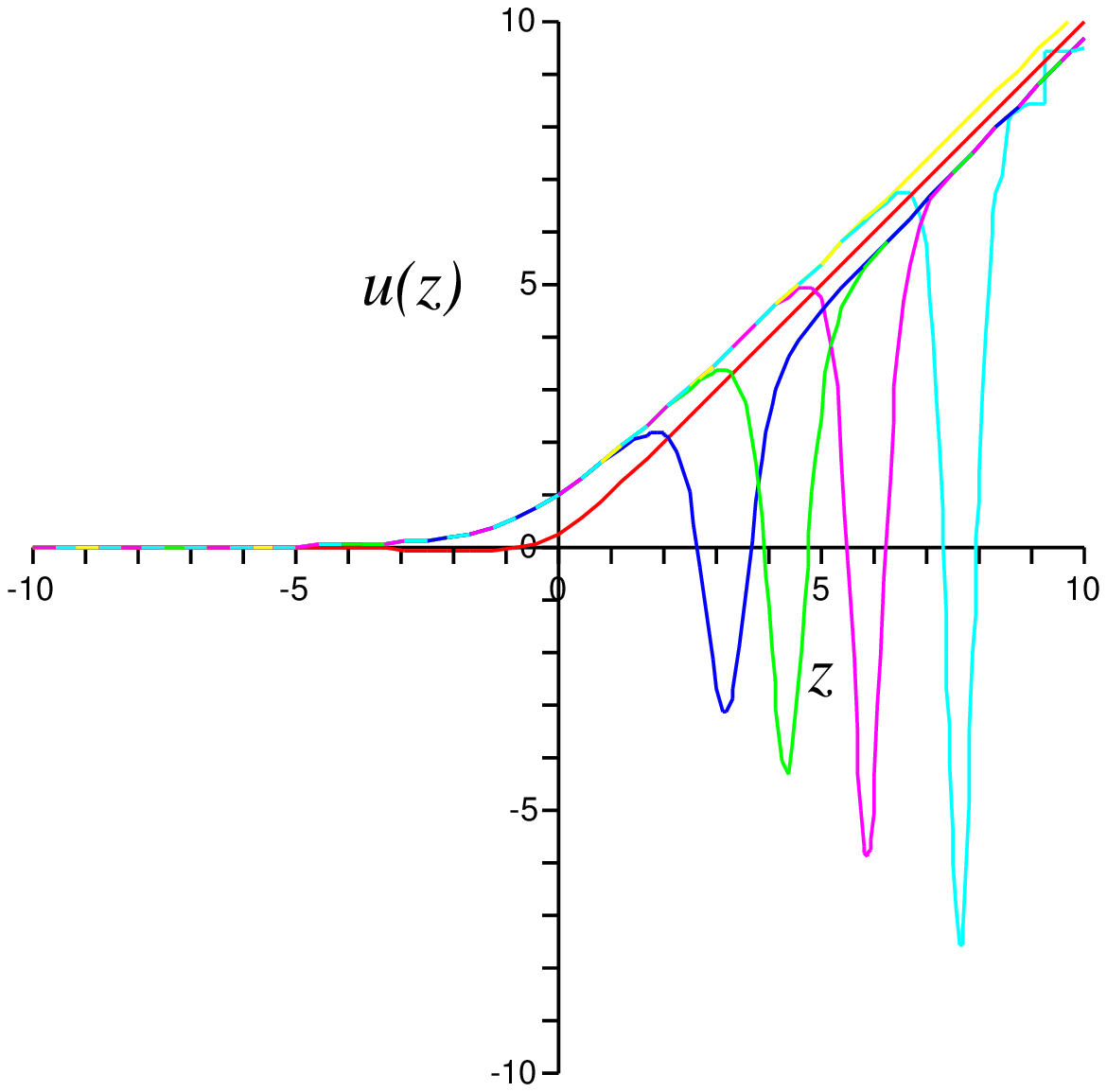}
\includegraphics[scale=0.55]{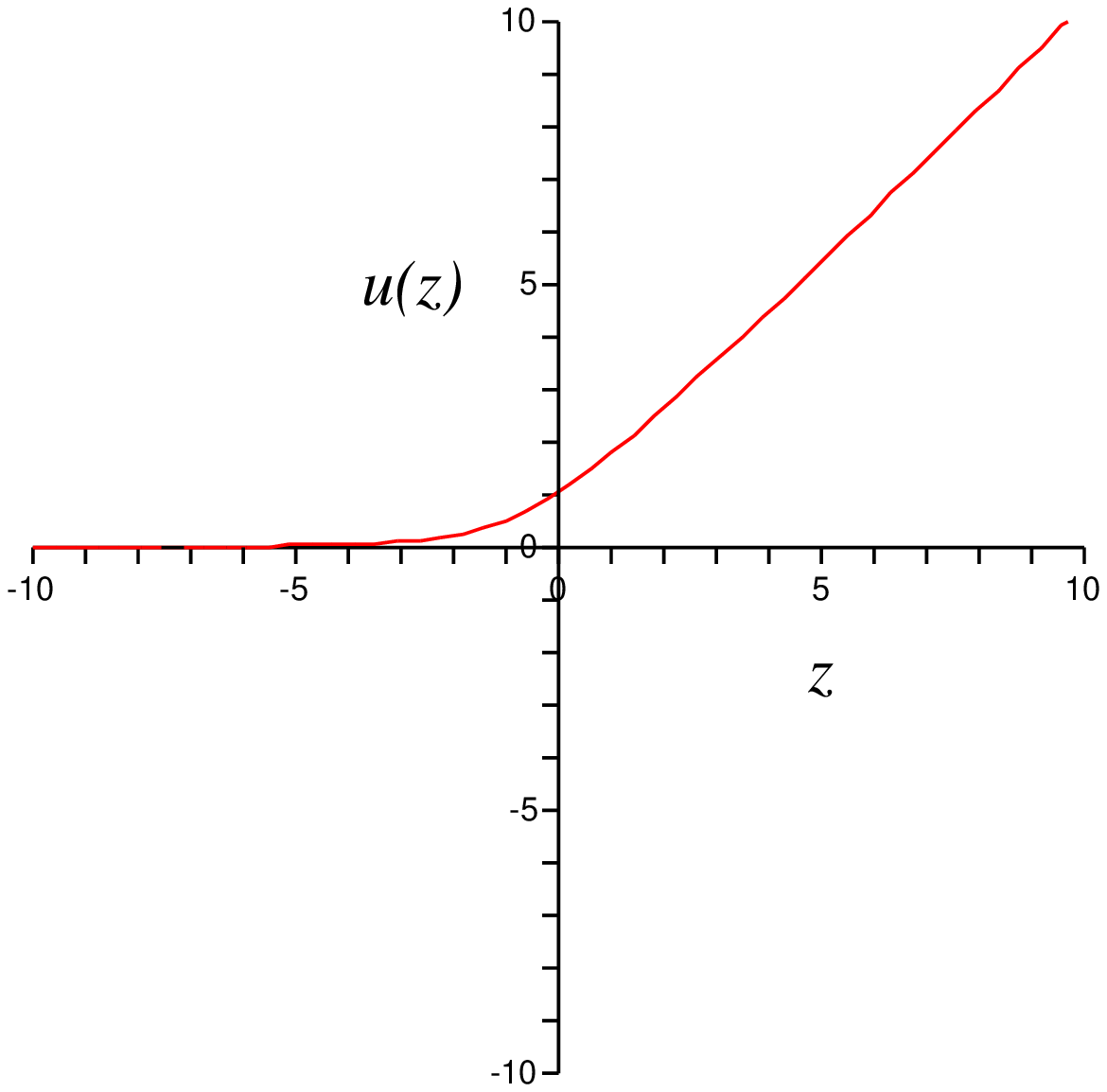}
\end{center}
\caption{\small  Numerical solutions to equation~\reef{eq:nonpert} for $u(z)$ in the case $k=1$. The set of curves on the left shows $u(z)$ for a range of negative values of  the parameter $\Gamma$, closely  approaching the value $-1$ from above. The curve with the lowest well (very close to $\Gamma=-1$) is beginning to show signs of numerical error to the right. The curve on the right is the case of $\Gamma=+1$, used for comparison in the text.}
\label{fig:gammamorenegplots}
\end{figure}

We carried this out with very interesting results. See
figure~\ref{fig:gammamorenegplots} for examples. We were able to use
this method to generate $u(z)$ for $\Gamma=-1+\epsilon$ where
$\epsilon$ could easily be taken as small as $\sim 10^{-10}$. The
potential well is observed to get more deep and narrow increasingly
rapidly as $\epsilon\to0$. In fact, we observe numerically that the
well runs to infinity along the line $u(z)=-z$, becoming infinitely
deep and narrow in the limit $\Gamma=-1$.
Most interestingly, note that since the deepening of the well is
accompanied by a narrowing, the system has a chance of preventing the
appearance of a dangerous bound state. We used the families of $u(z)$
functions found here as potentials for ${\cal H}$ and were indeed able
to verify with some confidence (to within acceptable error tolerance)
that there were no bound states present for any of the potentials.

So the case $\Gamma=-1$ is remarkably special, and in fact contains a
surprise. At first, it seems that it develops a pathology, since as
$\epsilon\to0$ the well was seen to grow infinitely deep, but at the
same time it became extremely narrow and, moreover, it moved off to
infinity to the right. So the solution may in fact be well--behaved.
In support of this, it is interesting to note the striking fact
(again, from studying families of curves such as those displayed in
figure~\ref{fig:gammamorenegplots}) that in the limit the $\Gamma=-1$
curve looks rather like the $\Gamma=+1$ curve!  The feature associated
to the well is confined to a highly localized narrow (and deep)
region, which has moved away. Everything that remains of the
$\Gamma=-1$ curve at finite $z$ falls (with considerable numerical
accuracy) on the $\Gamma=+1$ curve.

This numerical observation suggests the surprising result that the
functions $u_{\Gamma=-1}$ and $u_{\Gamma=+1}$ are actually identical,
and that the discrepancy from the piece at infinity actually
disappears in the limit.  We can prove this for all $k$. The
transformation~\reef{eq:Back-Explicit-Gam} can be used to generate
either of these functions starting with the case of $u_{\Gamma=0}$.
Putting this function and $\Gamma=0$ into the equation immediately
gives the result that $u_{\Gamma=-1}=u_{\Gamma=+1}$, since there is no
explicit appearance of $\Gamma$ in the transformation in order to
generate the difference between decreasing {\it vs} increasing~$\Gamma$.

Note that below $\Gamma=-1$ there is a transition to completely new
behaviour. There, the system no longer has a smooth solution with the
asymptotics given in equation~\reef{eq:largez}, and there must be a
singularity at {\it finite} $z$. The origin of the singularity can be
seen (for $k=1)$ by considering the form of the
transformation~\reef{eq:Back-Explicit-Gam}. The denominator of the
transformation is $4 \mathcal{R}^2$, where, for large positive $z$, we
have that
\begin{equation}
  \label{eq:rlimit}
 {\cal R}= u^k-z\simeq  \frac{\nu\Gamma}{z^{\frac{1}{2k}}}+\cdots\ ,
\end{equation}
and so for $\Gamma<0$, the $k=1$ solution must approach the line $u=z$
from below as $z\to+\infty$. Since $u(z)$ started out above this line
(at $z\to-\infty$), it must be that it crosses the line somewhere in
the interior. If this is the case, then somewhere in the interior, the
new solution $u_{\Gamma-1}$ must have a singularity, since the
denominator vanishes there. We expect similar arguments to hold for
the negative $\Gamma$ behaviour of other values of $k$.

So we have the very natural physical situation singling out $\Gamma$
as a positive integer, since any solution $u(z)$ with $-1<\Gamma<0$
can be used as a seed for our
transformation~\reef{eq:Back-Explicit-Gam} to generate a $u(z)$
solution with poles, which have a problematic interpretation.
Furthermore, any solution $u(z)$ for a positive fractional value of
$\Gamma$ can also be used, by applying the transformation successively,
to generate such solutions, and so they are in the same class as the
$-1<\Gamma<0$ solutions. The $u(z)$ for positive integer values of
$\Gamma$ are therefore special, in that they are not connected by our
transformation to any solutions with poles.

\section{Summary}

So we see that studying the string equation non--perturbatively using
a combination of exact and numerical methods reveals that the
solutions have non--perturbative sensitivity to the values of
$\Gamma$. In particular, it appears that for $\Gamma\geq-1$, the spectrum
of the model is continuous and bounded from below by zero, which is
consistent with perturbative expectations. This means, for example,
that the wavefunction of the extended brane\cite{Fateev:2000ik}
probe\cite{Maldacena:2004sn}, the wavefunction $\psi$ of ${\cal H}$,
thought of as a function $\psi(\lambda_s)$ has connected support.

For $\Gamma<-1$, the system does not seem to maintain a continuous,
bounded spectrum. $\Gamma =-1$ is a special value therefore, marking
the beginning of a new phase of the model. It is tempting to think of
this as a phase transition, but since we have every reason to think of
$\Gamma$ as a discrete parameter, this is probably not well
motivated.

We are able to reach all positive integer values of $\Gamma$ using our
special B\"acklund transformation, and we observe that the cases of
$\Gamma$ being a negative integer do not really exist, since
$\Gamma=-1$ is the same as $\Gamma=+1$, so successive application of
the transformation will not generate any new negative integer cases.
So the positive integer $\Gamma$ solutions constitute a very special set.
Meanwhile fractional values of $\Gamma$ can be connected by the
transformation to solutions with poles, and so seem less physical.  So
while we have not rigourously proven that the positive integers are
the only physical values selected by the string equation, it is
certainly encouraging to find that the equation has properties that
single out the positive integers as special.

This connects rather nicely to the rectangular matrix model
understanding of the role of $\Gamma$ mentioned in the first section:
If $M$ is ``row--like'' ($\Gamma > 0$), then the combination
$MM^\dagger$ will have $N+\Gamma$ eigenvalues, and precisely $\Gamma$
of them are zero. These $\Gamma$ extra zeros correspond to the
zero--velocity KdV solitons (see below).  On the other hand, if the
matrix $M$ is a ``column--like'' ($\Gamma < 0$) rectangular matrix
({\it i.e.,} more columns than rows) then the combination $MM^\dagger$
will have $N-|\Gamma|$ eigenvalues. So negative $\Gamma$ has no extra
zero eigenvalues, and does not correspond to having any background
branes.  In fact it seems to reduce the number of eigenvalues of the
matrix, but at large $N$ (the limit in which we have defined the
continuun model) this makes no difference.  It would be interesting to
determine what the negative $\Gamma$ regime may mean
physically\footnote{An attempt at interpreting  negative integer
  $\Gamma$ as corresponding to adding $\Gamma$ anti--branes cannot be
  correct (we thank Nathan Seiberg for a comment about this),
  particularly in view of our observation about the non--existence of
  a distinct branch of solutions for $u(z)$ in this regime.}.  This
regime presumably has a plethora of bound states, which would need to
given a physical interpretation.  Perhaps it is simply not possible to
do so, and so the development of singularities in $u(z)$ for
fractional $\Gamma$ would be indicative of a non--perturbative
sickness, reminiscent of that of the even $k$ bosonic
models\cite{Brezin:1990rb,Douglas:1990ve,Gross:1990vs}.

The observation that there are $\Gamma$ extra zero eigenvalues for
positive integer $\Gamma$ fits very well with the fact that our
specialization of the B\"acklund transformations creates and destroys
KdV solitons with spectral parameter ${\tilde \lambda}=0$. It is clear
that the localized D--branes of the minimal string theory should be
identified with these solitons. This direct connection of D--branes
with the solitons of the underlying integrable system is intriguing
and may well lead to interesting new physics in both string theory and
in the theory of integrable systems. It deserves further
investigation.

\section*{Acknowledgments}
JEC is supported by an EPSRC (UK) studentship at the University of
Durham. He thanks the Department of Physics and Astronomy at the
University of Southern California for hospitality during the course of
this project. JSP thanks the Department for Undergraduate research
support.  CVJ wishes to thank Stephan Haas for useful remarks, and the
organizers of the UCLA IPAM Conformal Field Theory Reunion conference,
Dec. 12th--17th 2004, where these results were presented.

\providecommand{\href}[2]{#2}\begingroup\raggedright\endgroup

\end{document}